\documentstyle[12pt]{article}
\begin{document}
\hbadness=10000
\hbadness=10000
\begin{titlepage}
\nopagebreak
\begin{flushright}
{\normalsize
DPSU-97-5\\
KEK Preprint 97-11\\
May, 1997}\\
\end{flushright}
\vspace{1cm}
\begin{center}

{\large \bf  Probing $\mu$-term Generation Mechanism\\
in String Models}

\vspace{1cm}

{\bf Yoshiharu Kawamura $^a$ 
\footnote[1]{e-mail: ykawamu@gipac.shinshu-u.ac.jp}}, 
{\bf Tatsuo Kobayashi $^b$
\footnote[2]{e-mail: kobayast@tanashi.kek.jp}}\\
and\\
{\bf Manabu Watanabe $^a$
\footnote[3]{e-mail: watanabe@azusa.shinshu-u.ac.jp}}

\vspace{1cm}
$^a$ Department of Physics, Shinshu University \\
   Matsumoto 390, Japan \\
and\\
$^b$ Institute of Particle and Nuclear Studies \\
   High Energy Accelerator Research Organization \\
   Midori-cho, Tanashi, Tokyo 188, Japan \\

\end{center}
\vspace{1cm}

\nopagebreak

\begin{abstract}
We give a generic method to select a realistic 
$\mu$-term generation mechanism 
based on the radiative electroweak symmetry breaking scenario
and study which type is hopeful within the framework of string theory.
We discuss effects of the moduli $F$-term condensation 
and $D$-term contribution to soft scalar masses.
\end{abstract}
\vfill
\end{titlepage}
\pagestyle{plain}
\newpage
\def\thefootnote{\fnsymbol{footnote}}

\section{Introduction}

One of important problems in the supersymmetric standard model (SUSY-SM)
is how a SUSY Higgs mass term, the $\mu$-term, is derived 
naturally within the framework of supergravity (SUGRA).
Because the $\mu$-term is not a soft SUSY breaking 
term and a natural order of such a mass $\mu$ is the gravitational 
scale $M$ in SUGRA\footnote{
Through this paper, we take $M=1$.}.
This problem is called the $\mu$-problem \cite{muprob}.
Some types of natural mechanisms of $\mu$-term generation
have been proposed, 
e.g. the case with K\"ahler potential including a Higgs mixing 
term \cite{mu1} and a superpotential including a suitably suppressed 
mass term \cite{mu4,muprob,mu3,mu2}.

In addition to the supersymmetric mass term, soft SUSY breaking mass 
terms also contribute to the Higgs mass terms.
Explication of SUSY breaking mechanism is an important 
and unsolved problem.
Superstring theory (SST) is only the known candidate for unified theory 
including gravity and is expected to give a natural solution
due to some non-perturbative effects.
Recently there have been various remarkable developments in studying
non-perturbative aspects of SST \cite{ST-duality}, but SUSY breaking
mechanism has not been fully understood yet.
For the present, we can derive formulae for soft SUSY breaking 
terms assuming the existence of nonperturbative superpotential
which induces SUSY breaking in string models \cite{IL,ST-soft}.
Further these soft terms can be parametrized simply using several
parameters in most cases.
For example, under the assumption that only the dilaton field $S$ 
and/or the overall moduli field $T$ trigger SUSY breaking by their
$F$-term condensations, soft scalar masses, gaugino masses
and $A$-parameters 
are written model-independently by the gravitino mass 
$m_{3/2}$ and the goldstino angle $\theta$ in the case with the vanishing 
vacuum energy \cite{BIM}.
We can extend such a parametrization into cases with several moduli 
fields \cite{multiT,BIMS}.
On the other hand, the $B$-parameter depends on the $\mu$-term 
generation mechanisms \cite{ST-soft,BIM,LN}.
Therefore $\mu$ and $B$ parameters are usually treated as
arbitrary parameters.
The more arbitrary parameters exist, the harder we analyze and the less 
we have a predictivity.
It is important to select $\mu$-term generation mechanism
in some way.

In this paper, we study several $\mu$-term 
generation mechanisms based on the radiative breaking scenario 
of electroweak symmetry breaking within the framework of string theory.
Examining the parameter space leading to successful electroweak 
symmetry breaking, we probe a realistic $\mu$-term generation mechanism.
We take into account $D$-term contributions to soft scalar masses 
\cite{D-term} in our analysis.\footnote{
Study on the Higgs sector has been done, e.g. within the 
framework of string models without $D$-term contribution \cite{Higgs-st}
and SUSY grand unified theory (SUSY-GUT) with $D$-term 
contributions \cite{GUTD}.}
The reason is as follows.
String models, in general, have extra gauge symmetries
including an anomalous $U(1)$ symmetry and extra massless matter fields 
other than the SUSY-SM ones.
We can reduce gauge group and massless spectrum 
through symmetry breaking along flat directions \cite{flat}.
As discussed in recent papers \cite{stringD}, sizable $D$-term 
contributions can appear through such flat direction breaking.

This paper is organized as follows.
In the next section, we give our basic assumptions,
formulae of $\mu$-term and $B$-term, and our strategy.
In section 3, we study several types of $\mu$-term generation mechanisms
based on the radiative breaking scenario and show
which type is hopeful.
Section 4 is devoted to conclusions and discussions.

\section{Method of Analysis}

\subsection{Basic assumptions}

We take the SUSY-SM with soft SUSY breaking parameters
expected to be derived as a low-energy theory 
from orbifold models through flat direction breaking as a starting point
of our analysis.
Let us first list our basic assumptions.
\begin{enumerate}
\item The SUSY is broken by the $F$-term condensations of dilaton
field $S$ and/or moduli fields $T_m$.

\item The vacuum energy $V_0$ vanishes, i.e., $V_0 = 0$,
at the tree level.

\item For the third family of the quark doublet $q_3$, the up-type singlet $t$ 
and the Higgs doublet $H_2$, their modular weights are $n_k = -1$.

\item The top Yukawa coupling $f_t$ does not depend on moduli fields, i.e.,
$\partial f_t/\partial T_m = 0$.

\item The modular weight of the Higgs doublet $H_1$ is $-1$ (or $-2$).

\item The product $H_1H_2$ of Higgs doublets is gauge invariant under
all gauge symmetries which are broken along a flat direction.

\item Effects of moduli-dependent threshold corrections for gauge couplings
and gaugino masses are negligibly small.

\item The Kac-Moody levels satisfy $k_\alpha = 1$.

\item The gaugino masses $M_{\alpha}$, $B$ and $\mu$ are all real.

\item The dependence of dilaton and moduli fields is small 
in the $\mu$-parameter, i.e.,
$\partial \mu/\partial S, \partial \mu/\partial T_m \ll 1$.
\end{enumerate}

The third and fourth assumptions can be justified from the fact that
the coupling $f_t$ is strong and allowed as a 
renormalized coupling in the untwisted sector.

Here we discuss the overall-moduli case ($T=T_1=T_2=T_3$) for simplicity.
Under the above assumptions, we can obtain the following formulae
of soft scalar masses $m^{(0)2}_k$, the $A$-parameter $A_t^{(0)}$ 
among $\tilde{q}_3$, $\tilde{t}$ and $H_2$, 
and the gaugino mass $M_{1/2}^{(0)}$,
\begin{eqnarray}
&~&m^{(0)2}_k = m_{3/2}^2 (1 + n_k \cos^2\theta) + d_k m_{3/2}^2,
\label{smass0}\\
&~&A_t^{(0)} = - \sqrt{3} m_{3/2} e^{-i\alpha_S} \sin\theta ,
\label{A0}\\
&~&M_{1/2}^{(0)} = \sqrt{3} m_{3/2} e^{-i\alpha_S} \sin\theta 
\label{gmass0}
\end{eqnarray}
where we use the following parameterization 
\begin{eqnarray}
&~& \langle (K^S_S)^{1/2} F^S \rangle
 = \sqrt{3} m_{3/2} e^{i\alpha_S} \sin\theta ,
\label{FS}\\
&~& \langle (K^{T}_{T})^{1/2} F^{T} \rangle
 = m_{3/2} e^{i\alpha_{T}} \cos\theta .
\label{FT}
\end{eqnarray}
Here $d_k m_{3/2}^2$ is a $D$-term contribution and we get the relation
$d_{H_1} m_{3/2}^2 + d_{H_2} m_{3/2}^2 = 0$ from the assumption 6 
and the fact
that its contribution is proportional to broken diagonal charges of
scalar field.
The natural order of $d_k$ is expected to be $O(1)$
from some explicit models \cite{stringD}.

In the limit of the moduli dominant SUSY breaking, we can not neglect
moduli-dependent threshold corrections, but we do not consider such a case 
for simplicity in this paper.

The top quark mass is given as $m_t(m_t)=f_t(m_t)
v|\sin \beta|/\sqrt{2}$\footnote{
The pole mass of top quark is related with the running mass as 
\begin{eqnarray}
m_t^{pole}=m_t(m_t)[1+{4\alpha_3(m_t) \over 3\pi}+O(\alpha_3^2)] .
\nonumber
\end{eqnarray}}
where $\tan \beta = \langle h_2 \rangle /\langle h_1 \rangle$
and $v^2 = \langle h_2 \rangle^2 + \langle h_1 \rangle^2$.
($h_i$'s are neutral components of $H_i$'s.)
We use the value $m_t=175$GeV as the top quark mass 
from the current experiments.
Through our analysis, we take the gauge coupling unification scale
$M_X = 1.7 \times 10^{16}$GeV as an energy scale 
where boundary conditions are imposed.
Hence it is supposed that the above formulae 
(\ref{smass0})--(\ref{gmass0}) hold at $M_X$.

By the use of renormalization group equations (RGEs) of the SUSY-SM 
\cite{RGE,manual},
soft scalar masses of $H_1$ and $H_2$ at the weak scale 
(we use the $Z$-boson mass $M_Z$) are given as,
\begin{eqnarray}
&~&m_{H_1}^2 = h m_{3/2}^2 + d m_{3/2}^2 ,
\label{smass1}\\
&~&m_{H_2}^2 = \bar{h} m_{3/2}^2 - d m_{3/2}^2 
\label{smass2}
\end{eqnarray}
where 
\begin{eqnarray}
&~&h = 2.56\sin^2\theta + (1+n_{H_1})\cos^2\theta,
\label{h}\\
&~&\bar{h} = 2.56\sin^2\theta - 3 I_{\Sigma} \sin^2\theta
\label{barh}
\end{eqnarray}
and $d \equiv d_{H_1} = -d_{H_2}$.
Here the second term in RHS of (\ref{barh}) represents
the effect of the top Yukawa coupling and $I_{\Sigma}$ is a function
of $\tan\beta$ (or the top Yukawa coupling).
The values of $\alpha_t(\equiv f_t^2/4\pi)$ and $I_{\Sigma}$ 
are given in Table 1.

\subsection{$\mu$-parameter and $B$-parameter}

Several types of solutions for the $\mu$-problem have been proposed
\cite{Munoz}.
Here we explain some of them briefly and we give formulae 
of the $B$-parameter
for each $\mu$-term generation mechanism.

($\mu$-1) The $\mu$-term $\mu_Z$ of $O(m_{3/2})$ 
appears after SUSY breaking in the 
case where a K\"ahler potential includes a term such as 
$Z H_1 H_2$ \cite{mu1}.
In this case we have 
\begin{eqnarray}
\mu_Z &=& m_{3/2} \langle Z \rangle 
- \langle {F}_T \rangle \langle Z^T \rangle ,
\label{mu1}\\
B_Z &=& 2 m_{3/2} + \langle {F}_T \rangle \left(\partial^T \log\mu_Z
-{n_{H_1}+n_{H_2} \over \langle T+T^* \rangle}\right)
\nonumber\\
&~& + {m_{3/2} \over \mu_Z}\langle {F}_T \rangle \langle Z^T \rangle .
\label{B1}
\end{eqnarray}
Hereafter we take $Z=1/(T+T^*)$ and then $\mu_Z$ and $B_Z$ 
are given as
\begin{eqnarray}
\mu_Z &=& m_{3/2} (1+e^{i\alpha_T}\cos\theta) ,
\label{mu1-ex}\\
B_Z &=& {m_{3/2} \over 1 + e^{i\alpha_T}\cos\theta} \{2-
\cos\theta(e^{-i\alpha_T}(1+n_{H_1}+n_{H_2}) - e^{i\alpha_T})
\nonumber\\
&~& - \cos^2\theta(2+n_{H_1}+n_{H_2})\} .
\label{B1-ex}
\end{eqnarray}

($\mu$-2) The $\mu$-term $\mu_\lambda$ of $O(m_{3/2})$ 
appears after SUSY breaking in the 
case where a superpotential $W$ includes
a term such as $\lambda \tilde{W} H_1 H_2$ \cite{mu2}.
Here $\tilde{W}$ is a superpotential which induces
SUSY breaking.
In this case we have 
\begin{eqnarray}
\mu_\lambda &=& \lambda m_{3/2} ,
\label{mu2}\\
B_\lambda &=& m_{3/2}\{2-e^{-i\alpha_T}
\cos\theta(n_{H_1}+n_{H_2}- \langle T+T^* \rangle
\langle \partial^T \log \lambda \rangle)\} .
\label{B2}
\end{eqnarray}

($\mu$-3) The $\mu$-term $\mu_\mu$ can be generated
through some non-perturbative effects such as gaugino condensation 
\cite{mu3} and it generally depends on the VEVs of $S$ and $T$.
In this case we have 
\begin{eqnarray}
\mu_\mu &=& \mu_\mu(S,T) ,
\label{mu3}\\
B_\mu &=& m_{3/2} \{-1 - \sqrt{3} e^{-i\alpha_S}
\sin\theta(1- \langle S+S^* \rangle 
\langle \partial_S \log \mu_\mu \rangle)
\nonumber\\
&~& - e^{-i\alpha_T}\cos\theta (3+n_{H_1}+n_{H_2}- 
\langle T+T^* \rangle
\langle \partial^T \log \mu_\mu \rangle)\} .
\label{B3}
\end{eqnarray}


($\mu$-4)  In the model with a singlet field $N$ which has a 
coupling $f_{N} N H_1 H_2$, the $\mu$-term appears
when the field $N$ develops its VEV at some lower energy
scale near $O(m_{3/2})$ \cite{mu4}.
As a result, the $B$-parameter can be generated from the $A$-term
$A_N f_{N} N H_1 H_2$.

There can be an admixture of several $\mu$-term
generation mechanisms and, in this case,
$\mu$ and $B$ parameters are given as
\begin{eqnarray}
\mu_{\rm Mix} &=& \sum_p \mu_p ,
\label{mu-mix}\\
B_{\rm Mix} &=& \sum_p \mu_p B_p / \sum_q \mu_q
\label{B-mix}
\end{eqnarray}
where the indices $p$ and $q$ run over all $\mu$-term
generation mechanisms.

Finally we discuss a multi-moduli case briefly.
$Z_{2n}$ and $Z_{2n} \times Z_M$ orbifold models \cite{Orb} 
have $U$-type of moduli fields corresponding to complex structures 
of orbifolds \cite{Umoduli}
and a mixing term in the K\"ahler potential as 
\begin{eqnarray}
{1 \over (T_3 +T_3^*)(U_3 +U_3^*)}(H_1H_2+ h.c.).
\label{multi-H}
\end{eqnarray}
In this case, the Higgs fields $H_1$ and $H_2$ belong to 
the untwisted sector.
We assume $F$-terms of $S$, $T_i$ $(i=1,2,3)$ and $U_3$ contribute 
SUSY breaking and these are parametrized by $m_{3/2}$, $\theta$, 
$\Theta_3$ and $\Theta_3'$ following Refs.\cite{BIM,multiT,BIMS}.
Then the soft scalar masses, the $\mu$-term and $B$-term are written 
at the tree level as \cite{BIMS,Higgs-st}
\begin{eqnarray}
&~& m_{H_1}^{(0)2}=m_{3/2}^2 (1-3\cos^2 \theta 
(\Theta^2_3+\Theta'^2_3))+d m_{3/2}^2, \\
&~& m_{H_2}^{(0)2}=m_{3/2}^2 (1-3\cos^2 \theta 
(\Theta^2_3+\Theta'^2_3))-d m_{3/2}^2, \\
&~& \mu_M =m_{3/2}(1+\sqrt 3\cos \theta (e^{i\alpha_{T_3}}\Theta_3+
e^{i\alpha_{T'_3}}\Theta'_3)), \\
&~& \mu_M B_M=2m_{3/2}^2(1+\sqrt 3\cos \theta e^{i\alpha_{T_3}}\Theta_3)
(1+\sqrt 3\cos \theta e^{i\alpha_{T'_3}}\Theta'_3) .
\label{multi-B}
\end{eqnarray}
It is easy to show that $B_M$ and $\mu_M$ reduce to $B_Z$ and $\mu_Z$,
respectively
if we set $\Theta_3 = 1/\sqrt{3}$ and ${\Theta}'_3 = 0$.

We give formulae of $\mu/m_{3/2}$ and $B/m_{3/2}$ in Table 2.
Here we choose $e^{-i\alpha_S} = 1$,
$e^{-i\alpha_{T_i}}=e^{i\alpha_{T_i}}=\pm1$
and $n_{H_1}=-1$ and use this choice in our analysis.
In general, the $B/m_{3/2}$ contains a small number of free
parameters compared with $\mu/m_{3/2}$ and so the analysis of
$B/m_{3/2}$ can be more predictable.

\subsection{Strategy}

In this subsection, we give an outline of our strategy to probe $\mu$-term
generation mechanism based on the radiative electroweak symmetry breaking
scenario.
The neutral fields $h_1$ and $h_2$ have the following 
potential \cite{Higgs}:
\begin{eqnarray}
V(h_1,h_2) &=& m_1^2h_1^2+m_2^2h_2^2+(\mu Bh_1h_2+h.c.) 
\nonumber \\
 &+& {1 \over 8}(g^2+g'^2)(h_1^2-h_2^2)^2 ,\\
m_1^2 &\equiv& m_{H_1}^2+\mu^2, \quad m_2^2~\equiv~m_{H_2}^2+\mu^2
\end{eqnarray}
where all parameters correspond to the values at $M_Z$.

The condition for the symmetry breaking is given as 
\begin{eqnarray}
m_1^2 m_2^2 < (\mu B)^2.
\label{SB}
\end{eqnarray}
The bounded from below (BFB) condition along 
the $D$-flat direction requires 
\begin{eqnarray}
m_1^2+m_2^2 >2 |\mu B|.
\label{BFB}
\end{eqnarray}
Further the conditions that minimizing the potential are given as
\begin{eqnarray}
&~& m_1^2+m_2^2 = -{2 \mu B \over \sin 2 \beta}, 
\label{mini1}\\
&~& m_1^2-m_2^2 = -\cos 2 \beta (M_Z^2+m_1^2+m_2^2) 
\label{mini2}
\end{eqnarray}
where we use the relation $M_Z^2 = {1 \over 4}(g^2+g'^2)v^2$.

By the use of stationary conditions (\ref{mini1}) and (\ref{mini2}),
$\mu$ and $B$ are expressed by other
parameters ($m_{3/2}$, $\cos\theta$, $\tan\beta$,$\cdots$), such that
\begin{eqnarray}
&~& {|\mu| \over m_{3/2}} = {1 \over \sqrt{2}}
\left({h-\bar{h}+2d \over -\cos 2\beta} - h - \bar{h} 
-({M_Z \over m_{3/2}})^2\right)^{1/2} ,
\label{mu}\\
&~& {|B| \over m_{3/2}} = {\sin 2\beta \over 2}{m_{3/2} \over |\mu|}
\left({h-\bar{h}+2d \over -\cos 2\beta} -({M_Z \over m_{3/2}})^2 \right) .
\label{B}
\end{eqnarray}
By the use of RGEs, we can obtain the $\mu$ and $B$ parameters 
at $M_X$ (they are denoted by $\mu_{\pm}^{(0)}$ and $B_{\pm}^{(0)}$,
respectively.)
as follows \cite{manual},
\begin{eqnarray}
&~& {\mu_{\pm}^{(0)} \over m_{3/2}} = \pm c_\mu {|\mu| \over m_{3/2}} ,
\label{mu0}\\
&~& c_\mu \equiv \left({\alpha_2(t_Z) \over \alpha^{(0)}}\right)^{3/2}
\left({\alpha_1(t_Z) \over \alpha^{(0)}}\right)^{1/22}
(1+6\alpha_t^{(0)} F(t_Z))^{1/4}
\label{cmu}
\end{eqnarray}
and
\begin{eqnarray}
&~& {B_{\pm}^{(0)} \over m_{3/2}} = \mp {|B| \over m_{3/2}} 
+ {\Delta B \over m_{3/2}} ,
\label{B0}\\
&~& \Delta B \equiv 3A_t^{(0)}{\alpha_t^{(0)} F(t_Z) \over
1 + 6\alpha_t^{(0)} F(t_Z)}
\nonumber \\
&~& + M_{1/2}^{(0)}\left\{t_Z(3\alpha_2(t_Z)
+{3 \over 5}\alpha_1(t_Z)) - {3\alpha_t^{(0)}(t_Z F'(t_Z)-F(t_Z)) \over
1 + 6\alpha_t^{(0)} F(t_Z)}\right\}
\label{DeltaB}
\end{eqnarray}
where 
\begin{eqnarray}
&~&\alpha^{(0)} \equiv {g^{(0)2} \over 4\pi}, ~~
\alpha_t^{(0)} \equiv {f_t^{(0)2} \over 4\pi},\\
&~&F(t_Z) \equiv \int_0^{t_Z} 
\left({\alpha_3(t) \over \alpha^{(0)}}\right)^{16/9}
\left({\alpha_2(t) \over \alpha^{(0)}}\right)^{-3}
\left({\alpha_1(t) \over \alpha^{(0)}}\right)^{-13/99}dt ,\\
&~&t_Z = (4\pi)^{-1} \log {M_X^2 \over M_Z^2} .
\label{tZ}
\end{eqnarray}
Here $g^{(0)}$ and $f_t^{(0)}$ are the gauge coupling
and top Yukawa coupling at $M_X$, respectively.

By comparing $B_{\pm}^{(0)}/m_{3/2}$ and 
$\mu_{\pm}^{(0)}/m_{3/2}$
derived from the
stationary conditions of radiative breaking with $B_p/m_{3/2}$
and $\mu_p/m_{3/2}$ ($p=Z,\lambda,\mu,M$) given in the last subsection, 
we can find 
allowable parameter regions for ($m_{3/2}$, $\cos\theta$, 
$\Theta_3$, $\Theta_3'$, $\tan\beta$, $d$, $\lambda$, $\mu_\mu$)
leading to successful electroweak symmetry breaking and
know which type of $\mu$-term generation mechanism is hopeful.
Our strategy, $\lq$bottom-up approach', is more generic and applicable 
than the usual one, $\lq$top-down approach',
where a realization of the radiative breaking scenario is examined
by defining the parameters $m_k^2$, $\mu$ and $B$ at $M_X$ and
checking if the quantities renormalized at $M_Z$ 
satisfy stationary conditions (\ref{mini1})--(\ref{mini2}).

\section{Which $\mu$-term is hopeful?}

In this section, we examine which type of $\mu$-term generation
is hopeful by taking the cases ($\mu$-1)--($\mu$-3) as examples
and comparing $B_{\pm}^{(0)}/m_{3/2}$ and 
$\mu_{\pm}^{(0)}/m_{3/2}$ with $B_p/m_{3/2}$
and $\mu_p/m_{3/2}$ determined by $\mu$-term generation mechanism.
We also give remarks of several extensions.
For our analysis, it is convenient to define the following functions,
\begin{eqnarray}
&~&{\cal B}_{\pm p} \equiv B_{\pm}^{(0)} - B_p ,
\label{calB}\\
&~&\mu_{\pm p} \equiv \mu_{\pm}^{(0)} - \mu_p .
\label{mupmx}
\end{eqnarray}
The ${\cal B}_{\pm p}$ and $\mu_{\pm p}$ 
should be satisfied the conditions
${\cal B}_{\pm p}=0$ and $\mu_{\pm p}=0$
by definition if we assume the radiative breaking of electroweak symmetry 
and the $p$-th type of $\mu$-term generation mechanism.

\subsection{Dilaton dominant case}

First we give allowed regions for $B_{\pm}^{(0)}/m_{3/2}$ and 
$\mu_{\pm}^{(0)}/m_{3/2}$
in the limit of dilaton dominant SUSY breaking
in Fig.1 and 2, respectively.
The ranges of $B_{-}^{(0)}/m_{3/2}$ and
$B_{+}^{(0)}/m_{3/2}$ are $-0.74 \sim 0.70$ and $-3.51 \sim -1.45$.
The ranges of $\mu_{+}^{(0)}/m_{3/2}$ and
$\mu_{-}^{(0)}/m_{3/2}$ are $3.20 \sim 4.98$ and $-4.98 \sim -3.20$.
Here we take $2 \le \tan\beta \le 10$\footnote{
In the case that we take a larger $\tan\beta$, it is necessary to
incorporate the contribution of Yukawa couplings other than
top quark neglected here.}, $d=0$ 
and $m_{3/2} \geq 50$GeV.
The value $\tan \beta =2$ ($10$) corresponds to 
$B_{-}^{(0)}/m_{3/2} =0.70$
 ($-0.74$), $B_{+}^{(0)}/m_{3/2}=-3.51$ ($-1.45$) and  
$\mu_{\pm}^{(0)}/m_{3/2}=\pm 4.98$ ($\pm 3.20$).
The inequality $m_{3/2} \geq 50$GeV is derived from the phenomenological
constraint $m_{\tilde{e}} \geq 84$GeV when $n_{e} = -1$.
The Fig.3 shows an allowable region of ($B_{\pm}^{(0)}/m_{3/2}$, $d$)
in the limit of $m_{3/2} \gg M_Z$.
In this way, we come to a conclusion that the $\mu$-term generation
mechanism can be realistic if the values of ${B}_{p}/m_{3/2}$ 
and $\mu_{p}/m_{3/2}$ hit the above ranges.

We study the first case ($\mu$-1).
It is shown that there is no region satisfying 
${\cal B}_{\pm Z}=0$ with $d=0$ from Fig.1 and $B_Z/m_{3/2} = 2$.
When $\tan\beta = 1.33$, there is a solution of ${\cal B}_{\pm Z}=0$.
However, in this case, it is not realistic since 
the top Yukawa coupling blows up below $M_X$.
This result is consistent with those in Ref.\cite{Higgs-st}.
Further we find no region satisfying $\mu_{\pm Z} = 0$ with $d=0$
from in Fig.2 and $\mu_Z/m_{3/2} = 1$.
As discussed in Ref.\cite{stringD}, $D$-term contribution can survive
even in the limit of dilaton dominant SUSY breaking if string model
contains an anomalous $U(1)$ symmetry which is cancelled by the 
Green-Schwarz mechanism \cite{GS}.
On the other hand, $D$-term contributions related to anomaly-free
symmetries vanish at the tree level in the limit of dilaton dominant 
SUSY breaking.
If we assume the existence of $D$-term contribution, 
there appears a region 
consistent with the condition ${\cal B}_{\pm Z}=0$. 
However it is a very narrow region with a relatively large positive value of 
$d$.
For example, the region with $\tan\beta \sim 2$ and $d \sim 7$ 
is allowed as given in Fig.3.
On the other hand, negative $D$-term contribution is needed 
to lower the value of $\mu_{+}^{(0)}/m_{3/2}$.
Hence it is impossible to realize these $\mu$-term generation even
with $D$-term contribution.

In the second case ($\mu$-2), we get the same result 
for $B$-parameter as the first one.
On $\mu$-parameter, we can estimate the value of $\lambda$
using the condition $\mu_{\pm \lambda}=0$.
Hence the radiative breaking scenario can be realized in the models 
with a relatively large positive $D$-term contribution, i.e., $d=O(10)$.

In the third case ($\mu$-3),
we have a solution even in the absence of $D$-term contribution.
The allowable region of ($m_{3/2}, \tan\beta$) is given in Fig.4.
The favorable value is $\tan\beta \le 2.8$ with $d=0$.
The introduction of $D$-term contribution yields $\tan\beta > 2.8$.
The $\mu$-parameter is treated as a free parameter 
as well as the second one since the origin is unknown.

\subsection{Dilaton and overall moduli case}

We consider the effect of overall moduli $F$-term condensation.
We have the same qualitative result for
$\cos\theta \ne 0$ as the dilaton dominant case.
That is, there is no allowed region in the first case,
but very narrow region exists with 
large positive $d$ in the second case and there exists 
an allowed region with natural values of ($m_{3/2}$,
$\tan\beta$, $d$) in the third case.
The above fact can be understood by the use of
the Eqs. (\ref{mu})--(\ref{tZ}) directly.
That is, both $\mu_{\pm}^{(0)}/m_{3/2}$ and $B_{\pm}^{(0)}/m_{3/2}$
are proportional to $|\sin\theta|$ in the absence of $D$-term
contribution and the limit $m_{3/2}^2 \gg M_Z^2$, 
and so they decrease as $\cos^2\theta$ increases.
Hence it is impossible to satisfy ${\cal B}_{\pm}^{(0)}=0$ 
in the first and second case with $d=0$.
The introduction of $d$ does not improve the situation drastically.

\subsection{Remarks of extension}

We discuss a multi-moduli case (\ref{multi-H})--(\ref{multi-B}).
For the case with $\Theta_3 = 1/\sqrt{3}$ and ${\Theta}'_3 = 0$,
we have no allowable regions for ${\cal B}_{\pm}^{(0)}=0$
and $\mu_{\pm}^{(0)}=0$ because this case corresponds
to the first case ($\mu$-1).
We can check that there exists an allowable region 
with $\tan\beta=2$,
$\Theta_3 = \Theta_3' = 0.38$ and  $\cos^2\theta = 0.99$.

We can carry out the case with $n_{H_1}=-2$.
For $\tan\beta=2$ and $d=0$, we find the following fact.
Compared with the case with $n_{H_1}=-1$,
the value of $B_{-}^{(0)}/m_{3/2}$ and $B_{+}^{(0)}/m_{3/2}$
decreases and increases, respectively and
the absolute value of $\mu_{\pm}^{(0)}/m_{3/2}$
decreases.
The difference between the values in the case with
$n_{H_1}=-1$ and $n_{H_1}=-2$ increases as $\cos^2\theta$ increases.
Hence the similar conclusion holds 
for the reality of the radiative breaking scenario
as the dilaton dominant case with $n_{H_1}=-1$.

In the case of an admixture of several $\mu$-term generation mechanisms,
we need a dominant contribution of the third mechanism ($\mu$-3)
to get an allowable region with natural values of ($m_{3/2}$,
$\cos\theta$, $\tan\theta$, $d$).

We discuss the case ($\mu$-4).
As we have an extra light singlet field $N$, 
the RG flows of $m_{H_1}^2$ and $m_{H_2}^2$ should be modified
owing to the effect of the Yukawa coupling $f_N$.
This case can be applied to a similar strategy discussed in the 
last section.
The difference is that $B_{\pm}^{(0)}$ receives 
RGE effects as an $A$-parameter above $M_Z$
and so we must compare the renormalized quantity
with not $B$-parameter but $A_N$ at $M_X$.
It is not discussed here further because the renormalized quantity
contains an unknown parameter $f_N$.

\section{Conclusions and Discussions}

We have given a generic method to select a realistic 
$\mu$-term generation mechanism 
based on the radiative electroweak symmetry breaking scenario
and studied which type is hopeful within the framework of string theory.
The $\mu$-term generated by some non-perturbative effects, i.e.,
($\mu$-3), can be hopeful
to realize the radiative symmetry breaking
scenario even in dilaton dominant supersymmetry breaking.
We have discussed effects of the moduli $F$-term condensation 
and $D$-term contribution to soft scalar masses.
In the case of overall moduli, we have the same qualitative result
as in the limit of dilaton dominant SUSY breaking,
that is, the first mechanism is impossible to realize the radiative
scenario, the second one is required to a large $D$-term contribution
of $O(10m_{3/2}^2)$ and the third one is hopeful.

Our method to select a realistic $\mu$-term generation mechanism
is so generic and powerful that we can apply it to
the case with an improvement of approximation and more
complex situations.
For example, the improvement by the incorporation
of 1-loop effective potential \cite{1loop}, 
the case with a large $\tan\beta$,
the case with large moduli-dominant threshold corrections for gaugino masses,
other assignments of modular weight for matter fields and
the modular dominant SUSY breaking case.\footnote{
In \cite{KKK}, $\mu$ and $B$ are studied in the multi-moduli case
with a large non-universality between Higgs masses.}
In the above situations, extra contributions $\Delta h$ and 
$\Delta\bar{h}$ are added to Eqs.(\ref{h}) and (\ref{barh}), respectively.
These studies have to be considered systematically to select
a realistic string model.

\section*{Acknowledgments}
The authors are grateful to S.~Khalil 
for useful discussions.

\newpage

\section*{Table Captions}

\renewcommand{\labelenumi}{Table~\arabic{enumi}}

\begin{enumerate}
\item The values of $\alpha_t$ and $I_{\Sigma}$.
Here we use $m_t = 175$GeV, i.e., $m_t(m_t) = 167.2$GeV
and $\alpha_t^{(0)} = \alpha_t(M_X)$.

\item The formulae of $\mu_p/m_{3/2}$ and $B_p/m_{3/2}$.
The second column shows the dilaton dominant SUSY breaking case
and the third one is the dilaton and overall moduli mixed case.
Here we take modular weights $n_{H_1}=n_{H_2}=-1$.
\end{enumerate}

\begin{center}
{\Large Table 1} 
\end{center}

\begin{center}
\begin{tabular}{|c|c|c|c|}\hline
$\tan\beta$ & $\alpha_t(m_t)$ & $\alpha_{t}^{(0)}$ 
& $I_{\Sigma}$ \\ \hline
2 & $9.16\times10^{-2}$ & $1.11\times10^{-1}$ & 4.179 \\
3 & $8.14\times10^{-2}$ & $4.03\times10^{-2}$ & 4.140 \\
4 & $7.79\times10^{-2}$ & $3.20\times10^{-2}$ & 4.104 \\
5 & $7.62\times10^{-2}$ & $2.90\times10^{-2}$ & 4.083 \\
10 & $7.40\times10^{-2}$ & $2.56\times10^{-2}$ & 4.051 \\ \hline
\end{tabular}
\end{center}

\begin{center}
{\Large Table 2} 
\end{center}

\begin{center}
\begin{tabular}{|c|c|c|}\hline
  & Dilaton dominant & Dilaton and Moduli \\ \hline 
 $\mu_Z/m_{3/2}$ & 1 & $1 \pm \cos\theta$ \\
 $\mu_\lambda/m_{3/2}$ & $\lambda$ & $\lambda$ \\
 $\mu_M/m_{3/2}$ & 1 & $1 \pm \sqrt{3}\cos\theta (\Theta_3+\Theta'_3)$ 
\\ \hline
 $B_Z/m_{3/2}$ & 2 & 2 \\
 $B_\lambda/m_{3/2}$ & 2 & $2 \pm 2\cos\theta$ \\
 $B_\mu/m_{3/2}$ & $-1 \mp \sqrt{3}$ & 
 $-1-\sqrt{3}\sin\theta\mp\cos\theta$ \\
 $B_M/m_{3/2}$ & 2 & $2{(1 \pm \sqrt{3}\cos\theta\Theta_3)
   (1 \pm \sqrt{3}\cos\theta\Theta'_3) \over 
   1 \pm \sqrt{3}\cos\theta (\Theta_3+\Theta'_3)}$\\
 $B_{\rm Mix}/m_{3/2}$ & ${2m_{3/2}(1+ \lambda) - \mu_\mu(1\pm\sqrt{3})
 \over m_{3/2}(1+ \lambda) + \mu_\mu}$ & 
 ${2m_{3/2}(1+ \lambda)(1\pm\cos\theta)- \mu_\mu(1+\sqrt{3}\sin\theta
 \mp\cos\theta)
 \over m_{3/2}(1\pm\cos\theta+ \lambda) + \mu_\mu}$
 \\ \hline
\end{tabular}
\end{center}

\newpage 

\section*{Figure Captions}

\renewcommand{\labelenumi}{Figure~\arabic{enumi}}

\begin{enumerate}
\item The values of $B_{\pm}^{(0)}/m_{3/2}$ versus $m_{3/2}$
with $\cos\theta = d = 0$.

\item The values of $\mu_{\pm}^{(0)}/m_{3/2}$ versus $\mu_{3/2}$
with $\cos\theta = d = 0$.

\item The values of $B_{\pm}^{(0)}/m_{3/2}$ versus $d$ with
$\cos\theta = 0$ in the limit of $m_{3/2} \gg M_Z$.

\item The values of $\tan\beta$ versus $m_{3/2}$ 
with $\cos\theta =  d = 0$.
\end{enumerate}

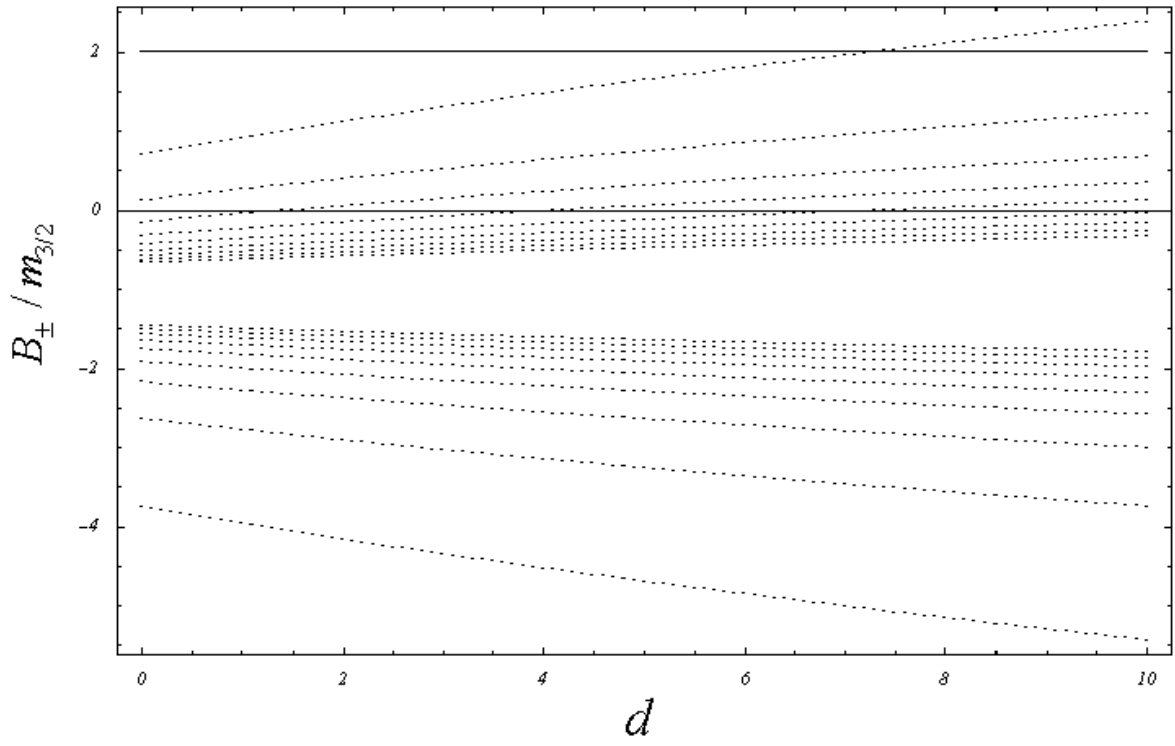
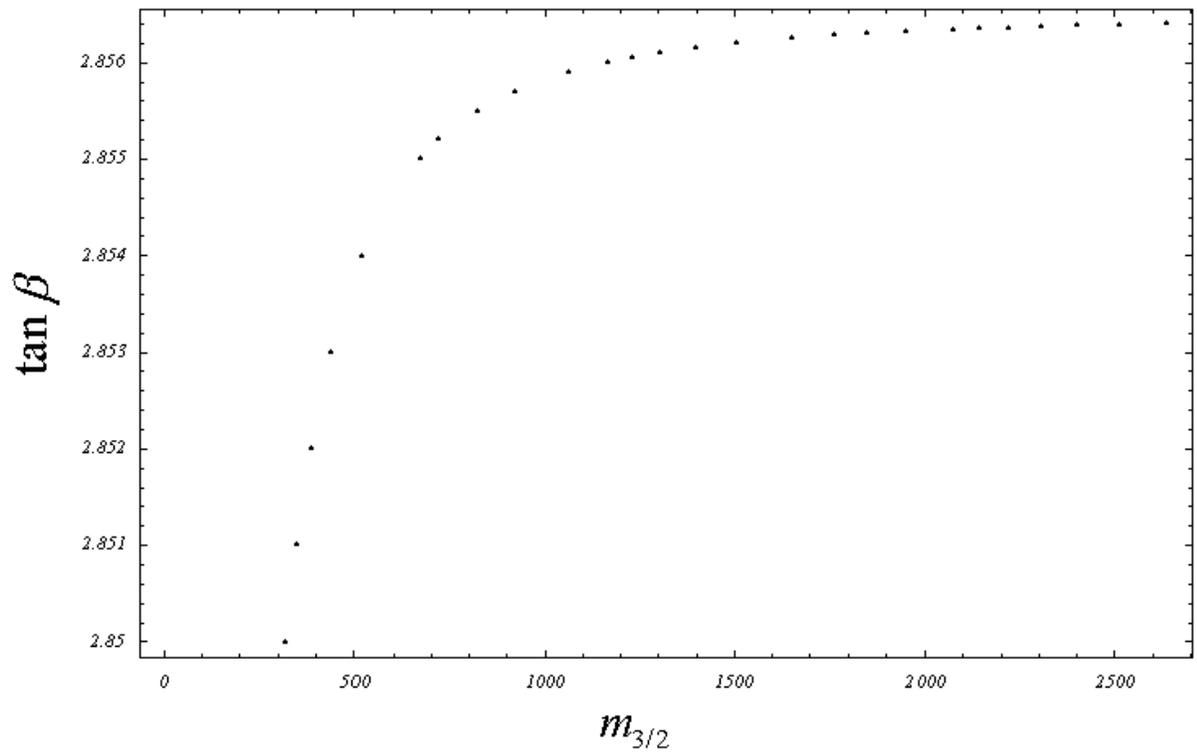
\end{document}